\documentclass[aps,prb,twocolumn,showpacs,showkeys,preprintnumbers,amsmath,amssymb]{revtex4-2}

\usepackage[utf8]{inputenc}
\usepackage{graphicx}
\graphicspath{{Figures/}}
\usepackage{hyperref}
\usepackage{units}
\usepackage{textgreek}
\usepackage{multirow}
\usepackage{rotating}
\usepackage{xcolor}
\usepackage{ulem}

\begin{document}
\title{Confinement Epitaxy of Large-Area Two-Dimensional Sn at the Graphene/SiC Interface}

\author{Zamin Mamiyev$^{1}$} \email{zamin.mamiyev@physik.tu-chemnitz.de}
\author{Niclas Tilgner$^{1,2}$}
\author{Narmina O. Balayeva$^{1,2}$}
\author{Dietrich R.T. Zahn$^{1,2}$}
\author{Thomas Seyller$^{1,2}$}
\author{Christoph Tegenkamp$^{1}$}
\address{$^{1}$Institute of Physics, Chemnitz University of Technology, Reichenhainer Str. 70, 09126 Chemnitz, Germany}
\address{$^{2}$Research Center for Materials, Architectures and Integration of Nanomembranes (MAIN), Rosenbergstr. 6, 09126 Chemnitz, Germany}

\begin{abstract}
Confinement epitaxy beneath graphene stabilizes exotic material phases by restricting vertical growth and altering lateral diffusion, conditions unattainable on bare substrates. However, achieving long-range interfacial order while maintaining high-quality graphene remains a significant challenge. Here, we demonstrate the synthesis of large-area quasi-free-standing monolayer graphene (QFMLG) via the intercalation of a two-dimensional (2D) Sn. While the triangular Sn(1$\times$1) interface exhibits a robust metallic band structure, the decoupled QFMLG maintains charge neutrality, confirmed by photoemission spectroscopy. Using high-resolution Raman spectroscopy and microscopy, we distinguish between direct intercalation and diffusion-driven expansion, identifying the latter as the critical pathway to superior QFMLG crystalline quality. Temperature-dependent analysis reveals dynamical structural coupling between the decoupled QFMLG and the Sn interface, providing a novel degree of freedom for strain engineering. Beyond uncovering the diffusion-driven mechanism, this work establishes metal intercalation as an effective strategy for tailoring durable graphene-metal heterostructures with tunable properties for next-generation quantum materials platforms.

\end{abstract}
{}

\maketitle
%************************************************************************

% Introduction
\section{Introduction}
%%%%%%%%%%%%%
In the field of advanced materials and nanotechnology, devising robust strategies to synthesize, stabilize, and functionalize 2D materials with atomic-level control remains a central challenge.\cite{Novoselov2016,Ares2022,Ferrari2015} Among the most scalable platforms, epitaxial graphene (EG) grown on SiC has matured into a benchmark for wafer-scale, uniform, high-quality 2D carbon, providing reproducible electronic and structural properties critical for both fundamental science and device integration.\cite{de2007epitaxial,Emtsev2009} Beyond serving as a model 2D conductor, epitaxial graphene is uniquely suited as a chemically inert, mechanically resilient “nanoreactor lid” that enables confinement epitaxy at buried interfaces \cite{Briggs2020}. In this architecture, the graphene overlayer kinetically suppresses three-dimensional clustering, whereas interface bonding and lattice registry guide the epitaxial ordering of intercalated metals and semimetals into atomically thin, laterally coherent films that are otherwise unstable on open surfaces. This interfacial “capsule” enables the formation of tailored electronic phases and superstructures across device-scale areas and provides a modular route to tune doping, strain, and hybridization without compromising the top graphene layer.
For example, Pb intercalation enables strain-mediated superstructures, inducing spin-orbit coupling into graphene \cite{Gruschwitz2025,Vera2024}, Sn and Si form a Mott-insulating phase in proximity to graphene \cite{Ghosal2025,Tilgner2025}, indium forms triangular lattices with robust quantum spin Hall insulator state \cite{Schmitt2024}, 2D Ga interface hosts superconductivity with higher T$_{C}$ than its bulk counterpart \cite{Wundrack2025}, or Ca intercalation induces superconductivity in graphene bilayers \cite{Ichinokura2016}. Particularly interesting is the Sn intercalation, which forms Mott insulating states with the correlated gap of $\sim$1.2 eV at one-third of monolayer coverage \cite{Ghosal2025}.  In contrast, the full monolayer coverage results in a metallic triangular (1$\times$1) lattice with decoupled charge-neutral QFMLG. \cite{Hayashi2017} The latter, possessing a high electronic density of states, interacts strongly and initiates Kekul\'e-O type bond density wave in the graphene with an energy gap of $\sim$90 meV. \cite{Thoai2025}
Recently, we found that a 2D Sn interfacial layer behaves as an embedded nanoantenna, forming a plasmonic gap mode together with an external plasmonic nanoantenna and greatly amplifies photon-graphene coupling \cite{Mamiyev2025a}. These features make Sn intercalation uniquely compelling, both from a fundamental perspective and for potential device concepts. Despite its promise, intercalation is not a simple material “switch”: its outcome is governed by multistep kinetics and thermodynamic pathways that determine the structure, uniformity, and emergent electronic phases \cite{Axdal1987}. 
\par Motivated by these considerations, we focus on Sn intercalation beneath epitaxial zero-layer graphene (ZLG) to elucidate its atomistic progression and quantify the resulting electronic and phononic modifications. Although ZLG shares the hexagonal honeycomb structure of free-standing graphene, its chemical bonding to the SiC substrate disrupts the $sp^{2}$ hybridization and breaks the lattice symmetry, preventing the formation of Dirac cones and rendering the layer non-metallic. \cite{Emtsev2008} Decoupling this layer via intercalation restores its metallic character, yielding a single-layer QFMLG with versatile properties mentioned above. Here, we demonstrate long-range ordered 2D Sn intercalation via diffusion-driven pathways, resulting in high-quality, charge-neutral QFMLG. Our findings reveal a distinct structural coupling between the graphene and the intercalated Sn layer, establishing a new route for controllable strain engineering in graphene-metal heterostructures.
\begin{figure*}[ht]
\centering
\includegraphics[width=0.9\textwidth]{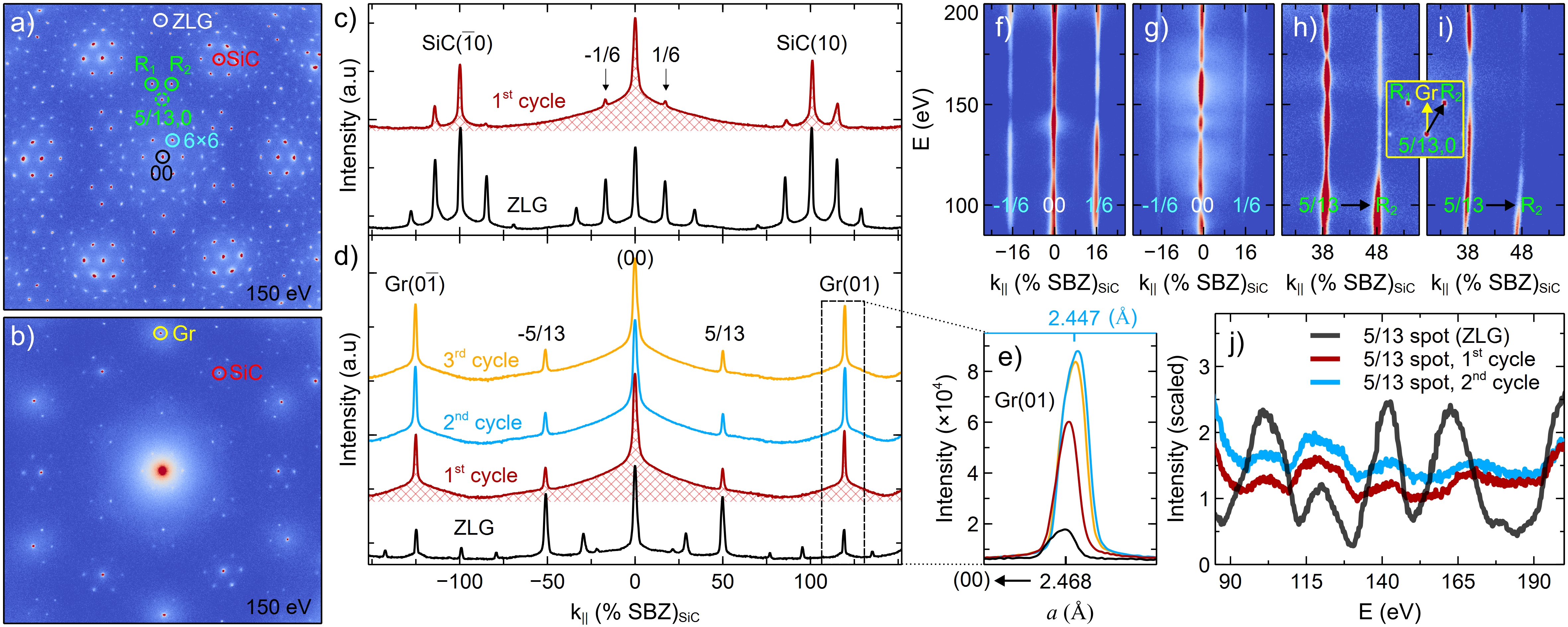}\\
\caption{\textit{In situ} study of the Sn intercalation and structural properties. a) SPA-LEED image for ZLG on SiC(0001). b) The same surface after Sn intercalation. The $R_{1}$ and $R_{2}$ in (a) denote the (6/13,-1/13) and (6/13,1/13) orders of the 6$\sqrt{3}$ periodicity.
c,d) High-resolution spot profiles along the SiC and Gr directions, respectively. The shaded areas mark the BSC. e) Close-up of the Gr(10) spot at different intercalation stages. The x-axis indicates the lattice constants ($a$) of ZLG (bottom, black) and QFMLG (top, blue), determined from the distance to the (00) diffraction spot (arrow).
f,g) Reciprocal space maps of the (00) spot along the SiC direction for pristine ZLG and after Sn intercalation, respectively. h,i) Reciprocal space maps for the (5/13,0) and (6/13,1/13) orders of the 6$\sqrt{3}$ periodicity (black arrow in the inset).
j) Intensity versus primary electron energy (E) plot extracted from (h,i). 
\label{LEED1}}
\end{figure*}
\section{Experimental methods}
\label{experiment}
Zero-layer graphene (ZLG) samples used for Sn intercalation and monolayer graphene (MLG) for reference were grown epitaxially by Si sublimation on  4H-SiC(0001) substrates as described in detail in Ref. \cite{Emtsev2009}. The samples were then transferred to the spot-profile analysis low-energy electron diffraction (SPA-LEED) chamber through the air and degassed at 850 K for several hours using direct-current heating. This process yields homogenous ZLG  structures with the characteristic ($\rm 6\sqrt{3}\times 6\sqrt{3}$)R30° reconstruction spots in SPA-LEED (Figure~\ref{LEED1}a).
A high-resolution SPA-LEED setup with a transfer width of 200 nm was used for surface studies to control and investigate the Sn intercalation process and interface structures \cite{Scheithauer1986,Mamiyev2021a}.
For intercalation experiments, Sn was deposited by electron-beam evaporation from a molybdenum crucible by keeping the sample at room temperature (RT) and applying a potential equivalent to the acceleration voltage of the evaporator to avoid sputtering effects from ionized Sn. \cite{Mamiyev2025} Temperatures were measured using an infrared pyrometer (Impac IGA) with an emissivity of 0.92. More details on Sn intercalation are reported in Ref.\cite{Mamiyev2022}.
Confocal Raman measurements were performed under ambient conditions using a Horiba Xplora Plus equipped with a DPSS continuous-wave (CW) laser source at different wavelengths with a 1200 mm$^{-1}$ grating, 3.0 mW laser power, and 1.6 cm$^{-1}$ spectral resolution. Temperature-dependent Raman measurements were conducted using a Linkam THMS600 stage under vacuum.
After transporting the samples through the air, photoelectron spectroscopy measurements were performed before and after degassing them at 870~K.
The chamber is equipped with a monochromatized SPECS XR50M Al-K\textalpha\ X-ray source and a monochromatized SPECS UVS 300 ultraviolet source that provides linear polarized He-I radiation. Photoelectrons were detected using a 2D CCD detector attached to a SPECS Phoibos 150 hemispherical analyzer. The total energy resolutions of the X-ray and angle-resolved photoelectron spectroscopy (XPS and ARPES) measurements were 100 meV and 40 meV, respectively.
%%%%%%%%%%%%%
\section{Results and Discussion}

\subsection{Electron diffraction studies}
\label{LEED}
Figure \ref{LEED1}a) shows a high-resolution SPA-LEED image of pristine ZLG on 4H-SiC(0001). Beyond the first-order SiC(1$\times$1) and ZLG($1\times1$) spots, the diffraction pattern reveals ($6\sqrt{3}\times6\sqrt{3}$)R30° ($6\sqrt{3}$ in the following) reconstruction spots of ZLG. These spots coincide with the moir\'e periodicity that arises from the superposition of graphene (Gr) and SiC lattices, imposed by a 30° rotational alignment and lattice mismatch. The appearance of sharp higher-order spots reflects the long-range order of the ZLG. Following Sn deposition at room temperature and subsequent annealing at 1075 K, the reconstruction spots are heavily suppressed (Figure \ref{LEED1}b-j), indicating the decoupling of the ZLG to a QFMLG. Moreover, the ZLG($1\times1$) spots are transformed into the Gr($1\times1$) spots, which appear brighter and are accompanied by an intense coherent background (also see Figure \ref{LEED1}b,d). This isotropic background, superimposed on the (00) spot and surrounding graphene reflexes, known as the bell-shaped component (BSC), is characteristic of free-standing graphene.\cite{Mamiyev2022} While its microscopic origin remains under debate, it may be attributed to low-energy flexural phonons that become active upon the decoupling of ZLG into a free-standing membrane, subsequently enhancing multiple-scattering events \cite{Fasolino2007,Lindsay2010} or to the uncertainty in the vertical momentum transfer.\cite{Chen2019} Its in-phase relationship with the Bragg component (Figure \ref{LEED1}g) demonstrates its link to the atomic lattice, rather than to defect-mediated or thermally induced scattering. \cite{Hoegen1999} 
\begin{figure}[ht]
\centering
\includegraphics[scale=0.5]{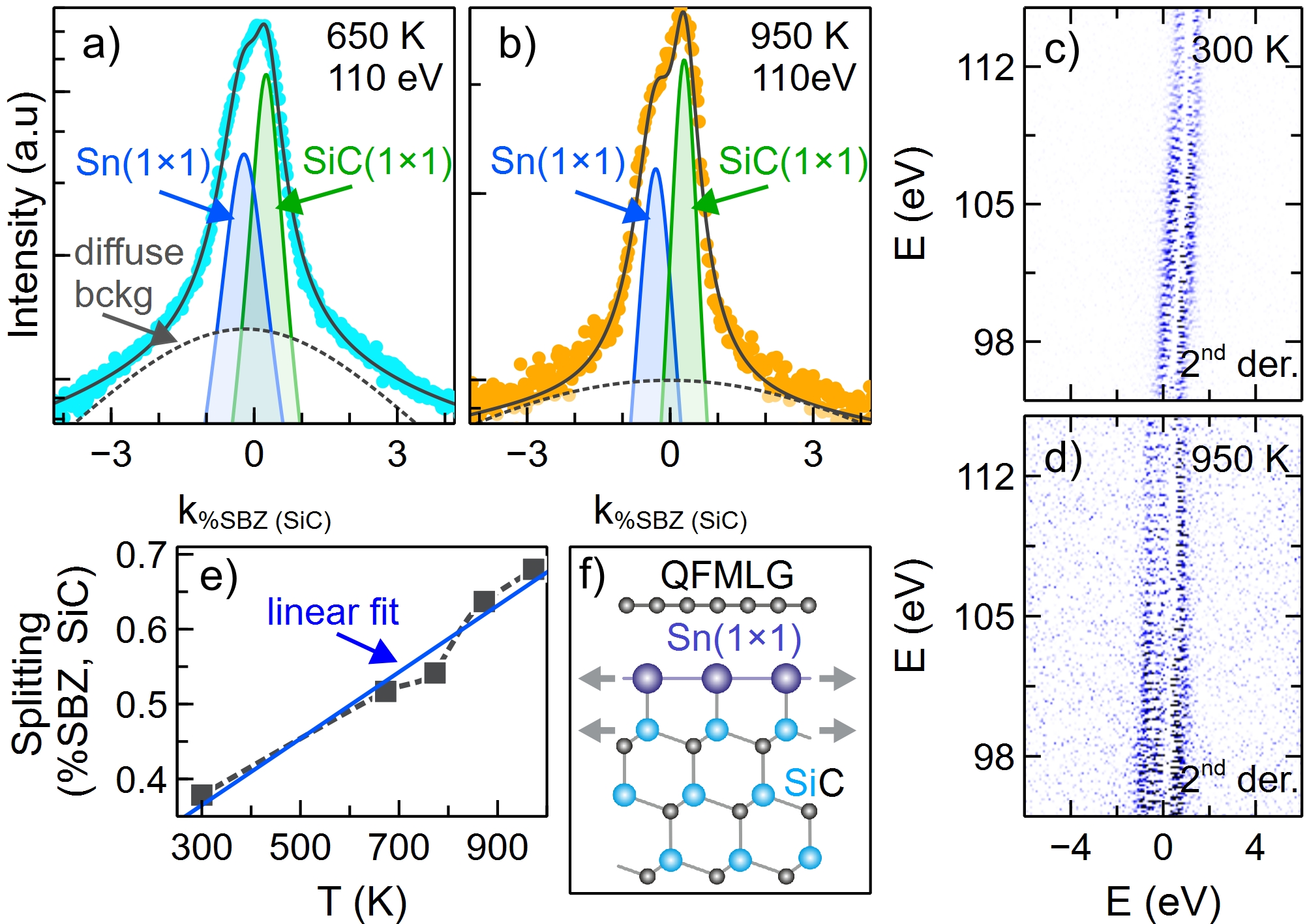}
\caption{a,b) Spot profiles of the SiC(10) spot at different temperatures. c,d) Reciprocal-space maps of the SiC(10) spot at 300 K and 950 K, shown as second derivatives for clarity. e) Lattice separation as a function of temperature. f) Side view of the Sn(1$\times$1) layer on SiC(0001); arrows denote the thermal lattice expansion.
\label{LEED2}}
\end{figure}
\par Assuming that the $6\sqrt{3}$ spots originate from the ZLG reconstruction and that the moiré-driven periodicity is quenched by the disruption of interfacial registry by Sn, their relative integrated intensity provides a quantitative measure of the surface fraction undergoing intercalation \cite{Mamiyev2022}. 
The analysis of the (5/13,0) and (6/13,$\pm$1/13)-spots of the $6\sqrt{3}$ periodicity reveals that their intensity is suppressed by $\sim$90\%, indicating only about 10\% of the surface remains non-intercalated after the first cycle. This observation is accompanied by an $\sim$82\% enhancement of the first-order graphene spots relative to the covalently bound ZLG($1\times1$) spots. Following a third annealing cycle at 1075 K for one hour, the degree of intercalation increases to $\sim$95\%. Notably, first-order SiC spots lose intensity, while the (00) specular spot gains intensity (Figure \ref{LEED1}b,c,f,g). This suggests the metallic nature of the Sn interface that, together with the topmost QFMLG, significantly screens the bulk SiC signal, while the higher scattering cross-section of Sn enhances the overall integrated specular intensity. 
\par The high-resolution spot profiles across the entire Brillouin zone reveal that the delaminated QFMLG shows a lattice constant $\sim$1\% smaller than the pristine ZLG (Figure \ref{LEED1}e). This observation aligns well with the previous comparison of the lattice constant of the ZLG and MLG systems using grazing-incidence x-ray diffraction. \cite{Schumann2014} It is evident that the QFMLG lattice significantly relaxes upon decoupling from the substrate, while the initial ZLG is subjected to tensile strain due to covalent C-Si bonding to the substrate.\cite{Ferralis2008,Schumann2014} Subsequently, the final lattice parameter at room temperature is governed by thermal expansion mismatch: as the system cools from the intercalation temperature, the negative thermal expansion coefficient (TEC) of graphene, contrasted with the positive TEC of the SiC substrate and the metallic Sn interface, induces residual compressive stress, consistent with the observed reduction in the lattice constant.\cite{Rohrl2008,Yoon2011a}
\par No additional superstructure spots are observed at this stage. The reciprocal-space maps of the (00) and reconstruction spots clustered around the 1/3 position (Figure \ref{LEED1}h–i, inset) likewise show no phase shift upon intercalation across primary electron energies of 85–200 eV (Figure \ref{LEED1}f–j). Together with the reduced intensity of the reconstruction spots, this indicates that although Sn does not adopt the symmetry of the ZLG, it remains coupled to the underlying substrate lattice potential. This suggests that the interfacial Sn conforms to the (1$\times$1) periodicity of the substrate (Figure \ref{LEED2}f). 
At 1 ML coverage, Sn atoms occupy all the ($1\times1$) sites of the SiC(0001) surface, forming a commensurate triangular lattice. 
\par To further elucidate the Sn($1\times1$) structure, we performed SPA-LEED measurements at elevated temperatures, focusing on the first-order SiC spots. The results show a clear separation of the Sn and SiC-related lattices (Figure~\ref{LEED2}a-d). While both Sn and SiC have positive TEC, obviously, the Sn lattice has a higher in-plane TEC than SiC, thereby enabling a clear separation of diffraction spots at elevated temperatures. The relative spot splitting at 950 K amounts to twice that at RT, which shows nearly linear behavior within the measured temperature range (Figure~\ref{LEED2}e).  
This corresponds to an expansion of the Sn lattice constant to 3.115 Å, yielding an effective linear TEC (\(\approx 10.5\times 10^{-6}\text{\ K}^{-1}\), via $\Delta\alpha/\alpha_{0}\Delta{T}$) nearly double that of the SiC substrate \cite{Li1986}. These results prove the hidden (1$\times$1) periodicity of the Sn interface, revealing that Sn is strongly compressed from its bulk lattice constant to conform to the SiC unit cell, which aligns well with recent calculations \cite{Visikovskiy2018}. The long-range ordering, therefore, requires interface corrugation (e.g., buckling) to compensate for the strong compressive strain. On the other hand, such interfacial roughnesses were found to play a role in reforming the graphene lattice into a Kekulé-O type bond density wave.\cite{Thoai2025,Mamiyev2025} Based on discussions above, the strain in the interfacial Sn periodicity and its possible effects on the graphene lattice are highly sensitive to the intercalation and postannealing temperatures.
\begin{figure}[ht]
\centering
\includegraphics[scale=0.39]{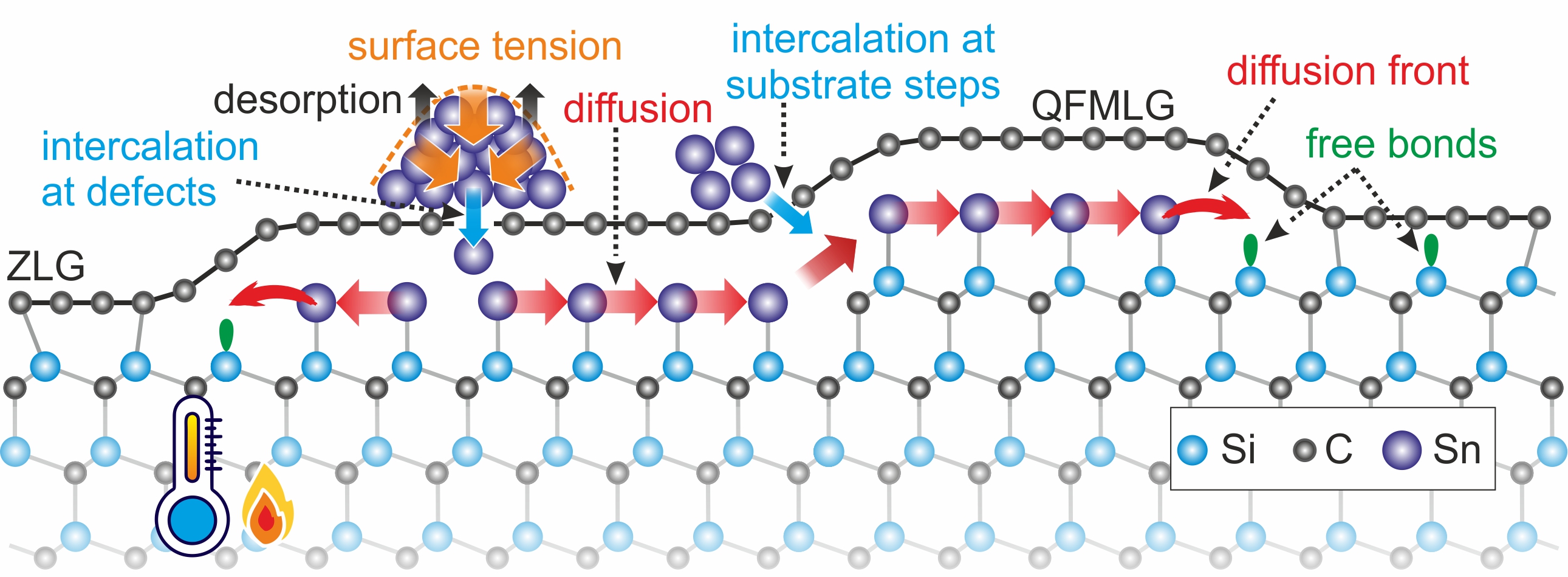}
\caption{Schematic view of the anticipated Sn intercalation process at elevated temperatures. 
\label{intmodel}}
\end{figure}
\par Although the exact conditions governing ($1\times1$) formation are not yet fully understood, this preferential structural template provides a densely packed interface that can be further rearranged or diluted into diverse configurations, enabling the modular tuning of the graphene doping level and interfacial interactions.\cite{Ghosal2025,Wundrack2025}
The anticipated scenario of the ($1\times1$) intercalation is shown in Figure~\ref{intmodel}. Intercalation typically starts at defects or step edges.\cite{Harling2025} 
The rapid lateral diffusion beneath the graphene, driven by chemical-potential gradients at the diffusion front and reinforced by vertical concentration gradients near entry points, produces a homogeneous layer governed by substrate registry. \cite{Thoai2025,Mamiyev2025}. 
Presumably, the balance between the intercalation and diffusion rate is the main reason to avoid multilayer and strained interface formation seen for other intercalants.\cite{Gruschwitz2025,Pompei2025}.
\begin{figure*}[ht]
\centering
\includegraphics[width=0.9\textwidth]{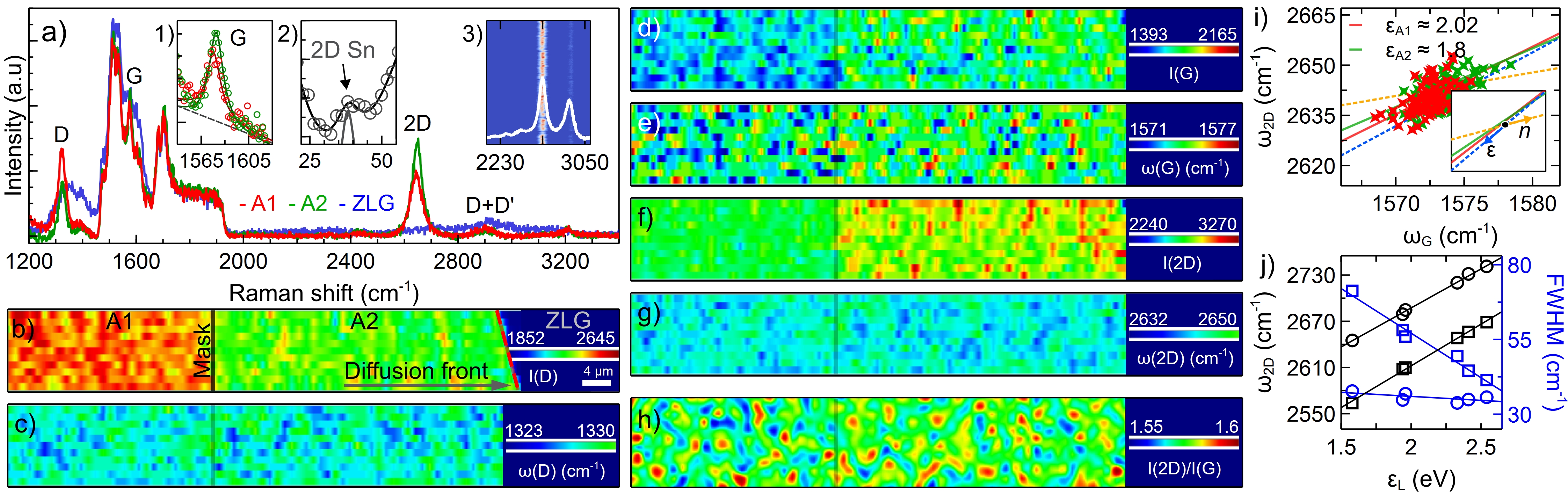}
\caption{Characterization of the samples via micro-Raman spectroscopy. a) Raman spectra of the intercalated QFMLG sample, recorded for direct Sn deposited (red, A1), diffusion-driven (green, A2), and nonintercalated (blue, ZLG) areas. The insets 1, 2, and 3 show zoom-ins of the graphene G band, low-frequency Sn band, and spatial mapping of the 2D band in the area A1 in (b).
b-g) Mapping of the Raman bands of the QFMLG across the shadow mask over an 80$\times$12 $\mu{m^{2}}$ area.
b) D band intensity, c) D band frequency, d) G band intensity, e) G band frequency, f) 2D band intensity, and g) 2D band frequency. h) I(2D)/I(G) ratio for the areas A1 and A2, relating lateral doping profiles. i) Strain and doping correlation in A1 and A2. The inset shows strain ($\epsilon$, blue) and doping ($\textit{n}$, orange) lines. 
j) Dispersions with excitation energy: frequencies (black) and width (blue) for the QFMLG/Sn (squares) and MLG (circles). The nonintercalated ZLG areas are masked in (b-h) for proper evaluation.
\label{Raman}}
\end{figure*}
\subsection{Raman spectroscopy studies}
Graphene shows characteristic Raman bands: the first-order G band (in-plane phonons at $\Gamma$), the second-order 2D band (double-resonant scattering at zone boundary), and the defect-activated D band.\cite{Ferrari2013} As the electron and phonon band structures of graphene are highly intertwined, these bands are sensitive to strain, doping, and electron-phonon coupling (EPC) \cite{Ferrari2013, Malard2009}. Raman results confirm the successful Sn intercalation, revealing the graphene G and 2D bands at $\sim$1575 and $\sim$2640 cm$^{-1}$, respectively (Figure~\ref{Raman} a). The red shifts relative to the MLG are primarily due to the charge neutrality of the QFMLG on the Sn interface, which resembles that of the freestanding single-layer graphene. \cite{Ferrari2013} The mismatch between the negative TEC of graphene and the positive TEC of the SiC/Sn interface induces frozen-in tensile strain upon cooling from the intercalation temperature, which, together with the vertical corrugation of the Sn($1\times1$) interface, leads to further phonon softening. The D band around 1330 cm$^{-1}$ is associated with defects in the graphene lattice and with finite patch sizes. \cite{Lazzeri2006,Ferrari2007} 
The Raman spectrum of QFMLG/Sn reveals a low-frequency phonon band at $\sim$40 cm$^{-1}$, which was previously assigned to the intercalated metallic Sn layer, confirming the long-range ordering and robustness of the interface in air. \cite{Turker2023} The 2D band measured across the sample (Figure~\ref{Raman}a, inset 3) shows a uniform frequency and width distribution, confirming the overall homogeneity of the decoupled QFMLG. Moreover, its symmetric line shape is well-described by a single Lorentzian, indicating a single sheet of QFMLG. 
\par Next, we employ micro-Raman imaging to probe the lateral transition from the directly Sn-deposited and intercalated area (A1) to the diffusion-mediated intercalated area (A2) below the shadow mask. 
It is obvious that Sn diffuses under the shadow mask to an area of $\textgreater$40 $\mathrm{\mu m}$ (Figure~\ref{Raman}b-g).  
The spatial variation of the D band intensity (Figure~\ref{Raman}b) shows that A2 exhibits a significantly lower defect density than A1, indicating higher-quality QFMLG. Since possible sputtering during electron-beam evaporation is largely suppressed by substrate biasing \cite{Mamiyev2025}, the observed defects are mainly attributed to
intercalation processes and finite QFMLG patch boundaries.
The approximately twofold lower D band intensity under the shadow mask strongly rules out dominant surface diffusion and demonstrates that Sn transport occurs
primarily at the interface.
A quantitative assessment of the defect density was obtained from the $I_\text{D}/I_\text{G}$ intensity ratio.\cite{Ferrari2007} 
At an excitation wavelength of 532~nm, this ratio is found to be 1.9 for A1 and 0.85 for A2, which fall into the regime typically associated with grain boundary or edge-type defects.\cite{Eckmann2012}
The defect density ($n_{D}$) and the mean distance between defects ($L_{D}$) were quantified using the Raman model by Lucchese \textit{et al}.\cite{Lucchese2010}, based on the $I_\text{D}/I_\text{G}$ ratio. The distance between defects is obtained from
$L_D^2=\frac{4.3\times10^{3}}{\epsilon_L^{4}}\left(\frac{I(D)}{I(G)}\right)^{-1}$, where $L_{D}$ is in nm and $\epsilon_{L}$ is the excitation energy in eV. The corresponding defect density (cm$^{-2}$) is then given by $n_D=\frac{10^{14}}{\pi L_D^2}$, the prefactor (in $nm^{2}eV^{4}$) is an empirical calibration constant according to Ref.\cite{Canccado2011}. For an excitation energy of 2.33 eV, we find $L_{D}\approx8.5$ nm and $n_{D}\approx4.1\times10^{11}$ cm$^{-2}$ for A1, which improves to $L_{D}\approx13.5$ nm and $n_{D}\approx1.8\times10^{11}$ cm$^{-2}$ for A2, values consistent with earlier reports.\cite{Lucchese2010,Canccado2011} 
These results suggest that intercalation at elevated temperatures induces defects in the graphene, which can be significantly reduced by lateral-diffusion-mediated intercalation. 
\par In accordance with the discussions above, the more intense G and 2D bands in A2 also support its higher quality (Figure~\ref{Raman}d,f).
Despite variations in intensity, the similar frequency distributions of the D, G, and 2D bands across both regions (Figure~\ref{Raman} c,e,g) suggest comparable electronic properties.
The $I_\text{2D}/I_\text{G}$ ratio (Figure~\ref{Raman}h), which reflects changes in doping, since increasing carrier density sharpens the G band via Pauli blocking and weakens the 2D band due to enhanced EPC, also indicates a uniform carrier concentration across the regions \cite{Zhao2011,Ferrari2013}.
The G and 2D bands were fitted with single Lorentzian profiles, and the resulting frequency correlation is shown in the strain–doping plot (Figure~\ref{Raman}i).
Using undoped, unstrained graphene as a reference ($\omega_\text{G}=1581.5$ $\mathrm{cm^{-1}}$ $\omega_\text{2D}=2677$ $\mathrm{cm^{-1}}$), strain and doping lines with slopes of 2.21 and 0.7 are plotted in Figure~\ref{Raman} i) \cite{Lee2012,Mueller2017}, confirming that tensile strain governs both the overall redshift and spatial frequency variations. The absence of G or 2D band splittings indicates biaxial strain \cite{Mohiuddin2009,Yoon2011}, consistent with previous STM observations \cite{Mamiyev2025}. 
The strain $\epsilon$ was quantified using a linear elastic model for biaxial strain: $\epsilon=-\frac{\Delta\omega}{\omega_{0}(1-v)\gamma}$, where $\Delta\omega$ is the measured frequency shift, $\omega_{0}$ is the reference frequency for undoped, unstrained graphene, $v$ is the Poisson ratio of graphene, and $\gamma_{0}$ is the Gr\"uneisen parameter. \cite{Mohiuddin2009,Yoon2011} Using $\gamma_\text{2D} = 2.7$ and $v=0.18$ \cite{Mohiuddin2009,Cheng2011}, the strain was estimated to be $\sim$0.4\%, where a similar value was also found from G band shifts ($\gamma_\text{G} = 1.8$). 
The results show that the frequency shifts in both regions are dominated by tensile strain, which is slightly weaker in A2. The moderately higher tensile strain in A1 is attributed to interface corrugation, whereas a smoother interface is expected for diffusion-driven intercalation \cite{Mamiyev2025}.
The tensile strain discussed here must not be confused with the global lattice contraction observed upon structural relaxation of the graphene membrane from its initially covalently bonded ZLG state, as analyzed by SPA-LEED, above. 
\par Figure \ref{Raman}j) compares the dispersion of the 2D band frequency and full width at half maximum (FWHM) for QFMLG/Sn and MLG, measured at varying excitation energies ($\epsilon_L$). The 2D band exhibits a linear dispersion with $\epsilon_{L}$, a hallmark of the double-resonant (DR) Raman process, while the first-order G band remains non-dispersive (not shown) \cite{Ferrari2013}.
Both samples exhibit a linear 2D band dispersion, fitted by $\omega_{\text{2D}}(\epsilon_{L}) = \omega_{0} + \beta\epsilon_{L}$, where $\beta$ is the dispersion slope and $\omega_{0}$ is the extrapolated intercept at $\epsilon_{L}=0$. QFMLG/Sn shows a steeper slope of $110~\mathrm{cm^{-1}\,eV^{-1}}$ compared to $100~\mathrm{cm^{-1}\,eV^{-1}}$ for MLG. Since ARPES measurements indicate a Fermi velocity ($v_{F} = 1.0\times10^{6}\,\text{m/s}$) for QFMLG comparable to that of MLG \cite{Ristein2012} (Figure \ref{arpes1} a), the observed difference in $\beta$ cannot be explained by $v_{F}$. Instead, it reflects modifications of the phonon dispersion and EPC, likely arising from strain, dielectric screening, and interfacial interactions in QFMLG/Sn. In intrinsically doped MLG, enhanced carrier screening and many-body effects partially suppress the Kohn anomaly near K(K$'$), leading to reduced phonon softening and a weaker 2D band dispersion \cite{Basko2009,Attaccalite2010}. 
The dispersion of the 2D band linewidth reinforces this picture (Figure \ref{Raman}j). In charge-neutral QFMLG, minimal electronic screening enhances the EPC near the Fermi energy; consequently, the DR process at low $\epsilon_{L}$ probes electronic states with shorter lifetimes. Higher $\epsilon_{L}$ shifts the resonant states toward regions of reduced scattering, resulting in a 2D linewidth narrowing of $\sim$30 $\text{cm}^{-1}$/eV. In contrast, doped MLG exhibits strong screening and Pauli blocking that suppress EPC-induced variations in lifetime, rendering the 2D linewidth $\epsilon_{L}$-independent. Since defects primarily affect absolute linewidths rather than phonon dispersion, they are not discussed further.\cite{Venezuela2011}
\begin{figure}[ht]
\centering
\includegraphics[scale=0.29]{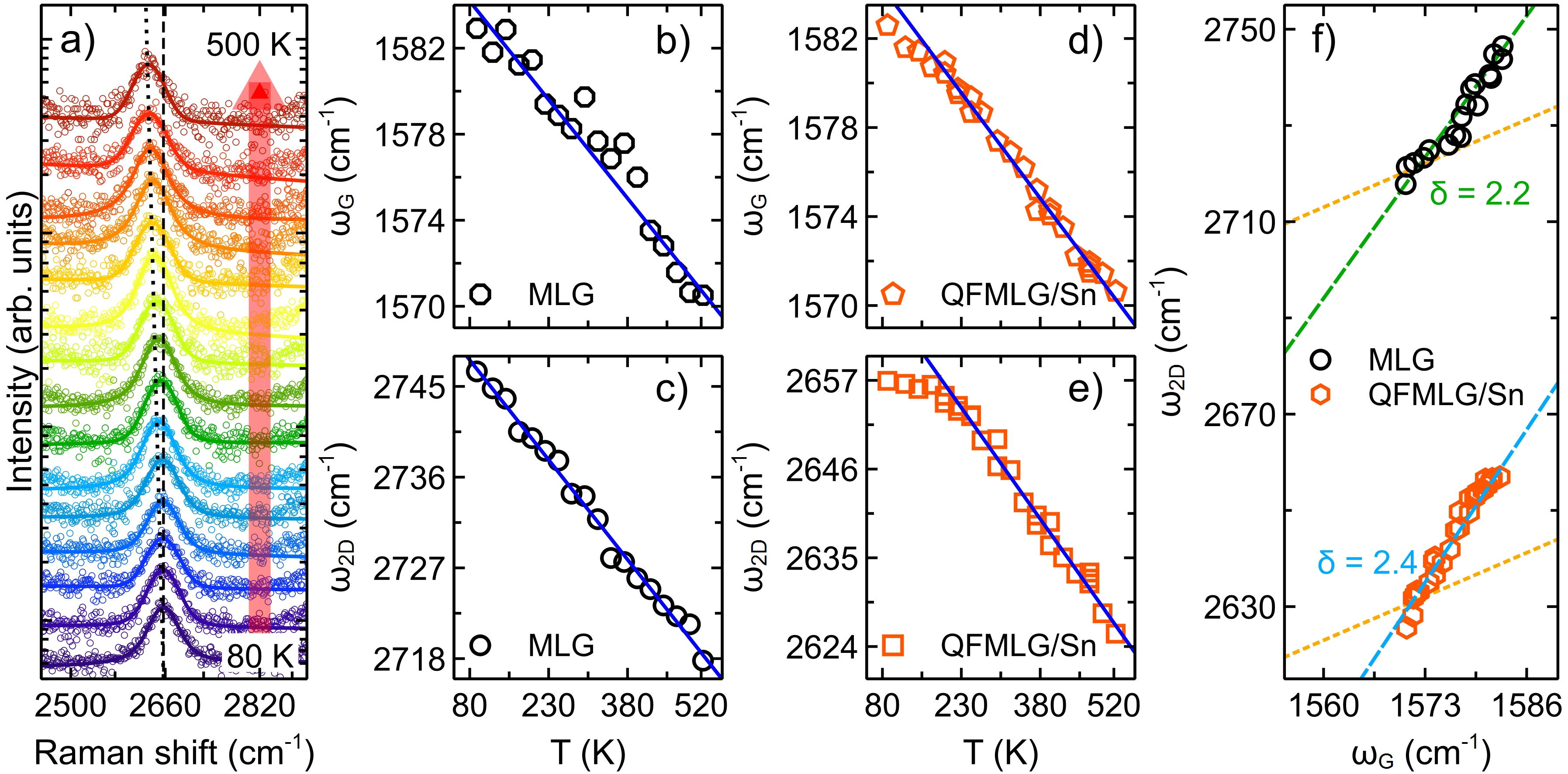}
\caption{a) Representative temperature-dependent Raman spectra of the 2D band for the QFMLG/Sn sample. b-e) TSRs of the G and 2D bands for the MLG and QFMLG/Sn samples; solid lines show linear fits. f) Correlation plot of the G and 2D band TSRs for both surfaces. Dashed (blue, green) and dotted (orange) lines indicate the strain and doping trends, respectively.
\label{Tdep}}
\end{figure}
\par Figure~\ref{Tdep} a-e) shows the thermal shift rates (TSRs) of the G and 2D bands for QFMLG/Sn compared to MLG in the temperature range of 80-520~K. For MLG, both G and 2D bands exhibit a linear red-shift with temperature ($T$), described by $\omega(T) = \omega_{0} + \chi T$, with slopes $\chi_{\text{2D}}^{\mathrm{MLG}} = -0.066~\mathrm{cm^{-1}K^{-1}}$ and $\chi_\text{G}^{\mathrm{MLG}} = -0.031~\mathrm{cm^{-1}K^{-1}}$, consistent with a strongly substrate-pinned regime where thermal expansion mismatch dominates the vibrational response.\cite{Ferralis2011}
In QFMLG, the G band TSR remains comparable ($\chi_\text{G}^{\mathrm{QFMLG}} = -0.032~\mathrm{cm^{-1}K^{-1}}$), indicating minimal modification of the zone-center optical phonon anharmonicity by intercalation. In contrast, Sn intercalation significantly enhances the 2D band TSR ($\chi_\text{2D}^{\mathrm{QFMLG}}=-0.092\text{\ cm}^{-1}\text{K}^{-1}$) while introducing a distinct plateau below 180 K (Figure~\ref{Tdep} e). The higher thermal response of QFMLG/Sn is attributed to the metallic Sn interface, which possesses a higher TEC than the underlying SiC, as observed by SPA-LEED, thereby acting as a thermal stress amplifier that exerts increased tensile strain on the graphene upon heating. The low-temperature invariance of the 2D peak position suggests a regime in which the intrinsic negative TEC of graphene compensates for the substrate expansion relative to the 80 K reference, potentially due to the reduced TEC of the Sn interface. \cite{Yoon2011a} Notably, the FWHM of both G and 2D bands remained temperature-invariant across all samples, confirming that the observed spectral shifts originate from thermal strain and lattice anharmonicity rather than thermally activated disorder or additional phonon scattering channels, thereby demonstrating excellent interfacial thermal stability. Using $T_{0} = 80$~K as a reference, we analyze the correlated TSR via $\Delta \omega_\text{2D} = \delta\,\Delta \omega_\text{G}$, where $\delta$ denotes the correlation slope and the temperature-induced shifts are defined as $\Delta \omega_{i}(T) = \omega_{i}(T) - \omega_{i}(T_{0})$ for $i \in \{\text{2D, G}\}$ (Figure \ref{Tdep} f).
We obtain \(\delta=2.2\) for MLG and \(\delta=2.4\) for QFMLG/Sn, indicating a strain-dominated correlation \cite{Das2008,Malard2009}. The slightly increased slope in QFMLG/Sn is consistent with the argument above regarding the role of the metallic Sn interface and aligns well with the SPA-LEED results. These findings demonstrate that Sn intercalation not only modulates the electronic environment but also fundamentally alters the interfacial coupling and thermal stability of the graphene-SiC interface, and introduces new possibilities towards dynamic strain engineering. 

\subsection{ARPES and XPS studies}
\begin{figure*}[ht]
\centering
\includegraphics[scale=0.9]{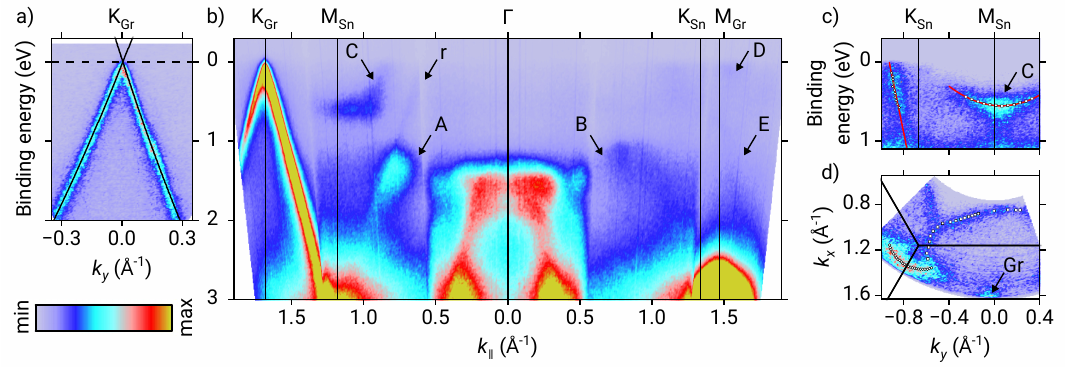}
\caption{Energy-momentum maps of the graphene-Sn heterostack.
a) Dirac cone of graphene measured perpendicular to the $\Gamma-K_{\mathrm{Gr}}$ direction of graphene. The straight lines depict the fitted dispersion with the nearest-neighbor tight-binding band structure of graphene. The horizontal dashed line marks the Dirac energy.
b) Overview map of the low-energy bands along the $\Gamma-K_{\mathrm{Gr}}$ and $\Gamma-M_{\mathrm{Gr}}$ directions of graphene. Capital letters indicate features of the Sn layer. Due to photoelectron diffraction, a graphene replica cone (“r”) is observed. c) Close-up of the Sn band structure along the K$_{\mathrm{Sn}}-\text{M}_{\mathrm{Sn}}$ direction. A linear and parabolic dispersing band is observed at the K$_{\mathrm{Sn}}$ and M$_{\mathrm{Sn}}$ point, respectively. These were evaluated using momentum and energy distribution curves (the maxima are depicted as circles and squares, respectively), and fitted with a straight line and a parabola, shown in red.
d) Fermi contour of the Sn layer. The straight lines indicate the boundaries of Sn’s 1$^\text{st}$ BZ. Two electron pockets around the K$_{\mathrm{Sn}}$ and M$_{\mathrm{Sn}}$ points are observed. Both contours were fitted using radial momentum distribution curves around the respective high-symmetry points. The MDC maxima are depicted as circles and squares, respectively.
All measurements were conducted at an approximate sample temperature of 20 K.}
\label{arpes1}
\end{figure*}
Further evidence of the ZLG decoupling is the appearance of a well-defined Dirac cone at the graphene K point (K$_{\mathrm{Gr}}$), as shown in Figure~\ref{arpes1}a). Fitting the maxima of momentum distribution curves (MDCs) with the tight-binding band structure of graphene (straight line) yields a Dirac energy of $E_{\text{D}} = -1 $~meV. Thus, the Sn-intercalated graphene layer is effectively charge-neutral, with a residual charge carrier concentration less than $1\times 10^{8}\,\text{cm}^{-2}$, aligning with the results from Raman spectroscopy.
\par As mentioned above and discussed further in this section, the confined Sn layer exhibits metallic behavior. This characteristic presumably explains the observed charge neutrality of the graphene. An important property of hexagonal SiC polytypes is their finite spontaneous polarization, which induces electron depletion in neighboring materials on the (0001) surface \cite{qteish1992electronic, Ristein2012}. However, in our case, no doping of graphene is observed, suggesting that the metallic Sn layer effectively screens the polarization field or compensates for electron depletion via charge transfer. Similar considerations were recently made for charge-neutral Pb/QFMLG \cite{Schadlich2023}. One should note that charge neutrality is observed for surfaces essentially free of residual Sn islands, which were found to induce mild $p$-type doping in graphene.
Figure~\ref{arpes1}b) shows an overview map along the high-symmetry directions $\Gamma-{\text{K}}_{\mathrm{Gr}}$ and $\Gamma-{\text{M}}_{\mathrm{Gr}}$. Beyond the intense contributions from graphene at K$_{\mathrm{Gr}}$ and M$_{\mathrm{Gr}}$, and the parabolic contribution from SiC at $\Gamma$, several distinct features (marked with capital letters) are observed, which are attributed to the 2D Sn layer. Specifically, there are fully occupied, dispersive states (labeled A and B) around $\Gamma$, a metallic band "C" around the M point of the SiC substrate (denoted M$_{\mathrm{Sn}}$), and a faint contribution "D" at the Fermi edge at the K point of SiC (denoted K$_{\mathrm{Sn}}$). These observations are consistent with density functional theory (DFT) calculations and previous experimental realizations of a 2D Sn layer on SiC(0001) with ($1\times1$) symmetry \cite{Hayashi2017,Visikovskiy2018}. This confirms the long-range ordering of the Sn layer, following the substrate periodicity, which is consistent with the SPA-LEED measurements above.
Additionally, a faint, linearly dispersing band "E" is observed near the K$_{\mathrm{Sn}}$ point. Unlike the other contributions, this state has not been reported before and is not explained by DFT calculations of the Sn($1\times1$) structure. Currently, the origin of this contribution remains unknown.

To gain a different perspective on the band structure of 2D Sn, Figure~\ref{arpes1}c) shows an energy-momentum map along the K$_{\mathrm{Sn}}-\text{M}_{\mathrm{Sn}}$ direction. Band “C”, which was previously observed to form a corner-like shape along the $\Gamma-\text{M}_{\mathrm{Sn}}$ direction, exhibits a parabolic shape along the perpendicular K$_{\mathrm{Sn}}-\text{M}_{\mathrm{Sn}}$ direction. The dispersion was fitted with a parabola, shown in red. Moreover, a linear dispersing band is visible near the K$_{\mathrm{Sn}}$ point. 
The linear and parabolic contributions observed in Figure~\ref{arpes1}c) are also evident in the Fermi surface depicted in Figure~\ref{arpes1}d). The intersection of these contributions with the Fermi level creates two electron pockets centered around the K$_{\mathrm{Sn}}$ and M$_{\mathrm{Sn}}$ points. As a visual aid, the maxima obtained from fitting radial MDCs are superimposed as circles and squares, respectively.
Note that the measurements presented in this section were conducted after an exposure to ambient conditions. However, the chemical composition of the sample surface was not altered in this process. Corroborating the Raman spectroscopy results, the confined Sn layer showed no signs of oxidation, as evidenced by the observed metallic band structure and Sn\,3d core-level spectrum (see below). This emphasizes the environmental stability provided by the protective graphene membrane, rendering confinement epitaxy a viable route for \textit{ex situ} characterization and application.
\begin{figure*}[ht]
\centering
\includegraphics[scale=0.9]{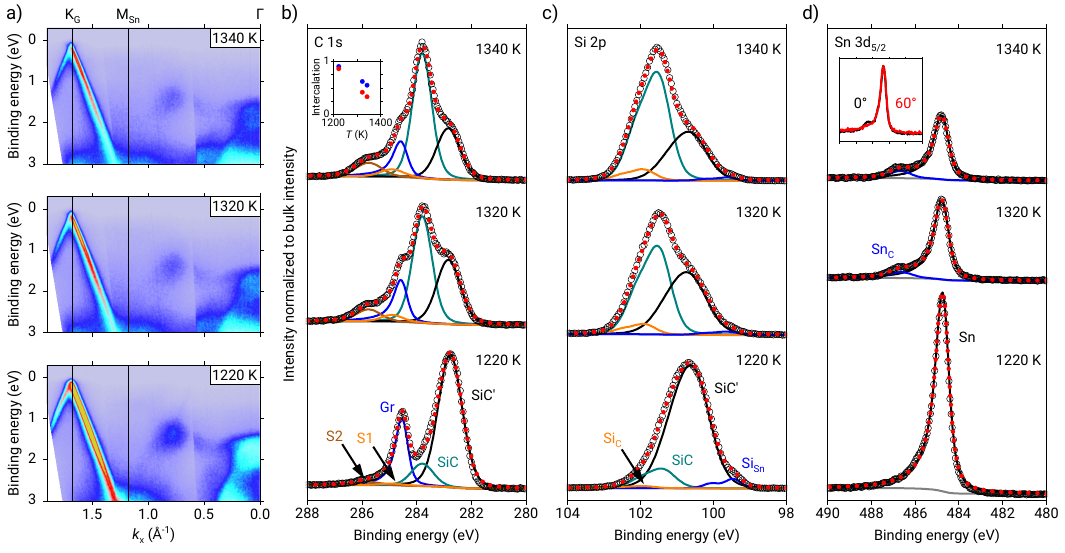}
\caption{Thermal stability of the graphene-Sn heterostack.
a) Energy-momentum maps along the $\Gamma-\text{K}_{\mathrm{Gr}}$ direction of graphene, obtained after annealing at the depicted temperatures. A decrease in the intensity of the graphene and Sn bands is observed with increasing temperature.
b–d) Corresponding core level spectra of the C\,1s b), Si\,2p c), and Sn\,3d$_{5/2}$ d) orbitals. Multiple components were used to fit the spectra. The sum of all components is shown as a red, dotted line. The inset of b) displays how relative intercalation changes with temperature. The blue (red) dots indicate the calculations using the graphene (SiC) intensities of the C\,1s XP spectra. The inset of d) shows the Sn\,3d$_{5/2}$ spectrum acquired at two different angles between the sample normal and analyzer. Refer to the text for discussion.
All measurements were conducted at room temperature.
\label{arpes2}}
\end{figure*}
The thermal stability of the QFMLG/Sn system was investigated by tracking its evolution under high-temperature annealing using ARPES and XPS. Significant changes were observed only above 1220 K. In energy-momentum maps recorded along the $\Gamma-\text{K}_{\mathrm{Gr}}$ direction (Figure~\ref{arpes2}a), both Sn- and graphene-related bands gradually lose intensity with increasing temperature, indicating progressive deintercalation. At the same time, the SiC valence band near \textGamma\ shifts to higher binding energies, which is consistent with the recovery of the ZLG. The ZLG is known to cause a weaker substrate band bending due to dangling bonds that pin the Fermi level \cite{Emtsev2008, Ristein2012}. No additional bands were observed upon high-temperature annealing.
Figures~\ref{arpes2}b-d) show the corresponding XPS results of the C\,1s, Si\,2p, and Sn\,3d$_{5/2}$ orbitals. The difference in substrate band bending is also reflected in the core levels by the two bulk components, SiC and SiC’, which correspond to SiC beneath the ZLG and the Sn-intercalated regions, respectively. QFMLG and ZLG contribute to the C\,1s spectra at higher binding energies. The graphene contribution was fitted with an asymmetric line shape that accounts for its metallic character. The ZLG parts were fitted with two components (S1 and S2) to account for the chemical shift between carbon atoms bound solely within the graphitic layer (S2) and carbon atoms coupled to the substrate (S1) \cite{Emtsev2008}. The Si\,2p spectra show two surface contributions, Si$_{\mathrm{Sn}}$ and Si$_{\mathrm{C}}$, next to the two bulk peaks. Si$_{\mathrm{Sn}}$ reflects the chemical shift due to the coupling of terminating Si atoms with intercalated Sn atoms. Si$_{\mathrm{C}}$ represents the Si atoms in nonintercalated regions connected to the ZLG. Up to an annealing temperature of 1220 K, the Sn\,3d$_{5/2}$ signal could be fitted with a single asymmetric peak, reflecting the metallic character of 2D Sn, as previously demonstrated by the low-energy band structure.

Significant changes are observed in the XP spectra upon annealing at temperatures greater than 1220 K. In general, the intensity of components associated with Sn-intercalated areas decreases, while the intensity of ZLG-related peaks increases. This reflects the desorption of Sn and the reemergence of ZLG through deintercalation.
The relative amount of intercalated regions can be quantified by the intensity ratios of the spectral components. The inset of Figure~\ref{arpes2}b) shows the decreasing trend of intercalation with increasing temperature. The blue dots represent calculations based on the intensity of the graphene component relative to the total intensity of the graphitic overlayer (i.e., graphene, S1, and S2). The red dots display the ratio of the SiC' component relative to the total bulk intensity. Note that the former calculation does not account for MLG growth during high-temperature annealing, which leads to an overestimation of the relative intercalation. The stepwise annealing process, conducted up to 1340 K, therefore decreased the initial $\approx$90\% intercalation to approximately one-third of the sample surface. This deintercalation temperature is relatively high, and the rate is significantly slower than previously reported.\cite{Mamiyev2022} We assign this difference to the multiple annealing cycles at the intercalation temperature (Figure~\ref{LEED1}d,e), which improves the long-range ordering and robustness of the Sn interface. While our previous work demonstrated that deintercalation of two-thirds of the Sn($1\times1$) interface coverage at $\sim$1320 K produces Sn($\sqrt{3}\times\sqrt{3}$)R30° periodicity with Mott-insulating behavior \cite{Ghosal2025,Mamiyev2024}, the present study shows that enhanced ordering suppresses deintercalation and reordering, and therefore must be considered when targeting phase transformations.
\par Furthermore, a new peak (Sn$_{\mathrm{C}}$) appears at higher binding energies in the Sn\,3d$_{5/2}$ spectrum, indicating a chemical bond to a more electronegative species. Since only carbon is available as such, the Sn atoms presumably become embedded in either the graphene layer or the substrate. Furthermore, the inset of Figure~\ref{arpes2}d) compares a Sn\,3d$_{5/2}$ spectrum with the Sn$_{\mathrm{C}}$ shoulder acquired under normal emission (black line) to a subsequent measurement under a 60° emission angle (red line). This comparison demonstrates that the Sn$_{\mathrm{C}}$/Sn ratio decreases with increasing emission angle, and thus, increasing surface sensitivity. These results support the interpretation that Sn is embedded in the substrate upon high-temperature annealing. The ratio of this shoulder to the remaining Sn amount shows that, after annealing at 1340 K, $\sim$10\% of the intercalated surface underwent such replacement of the Si topmost atoms with Sn, which may be interpreted as tin carbide. As such, a replacement reaction is highly unlikely on bare surfaces \cite{Kestle2000}; the graphene lid strongly modifies reaction kinetics, opening new ways of chemical engineering. On the other hand, this observation also emphasizes how cautiously the intercalation process must be taken, particularly when high temperatures are required.

\section{Summary and Conclusion}
In summary, we establish confinement epitaxy as an effective route to stabilize a long-range-ordered 2D metallic Sn layer at the graphene/SiC(0001) interface.
Sn intercalation decouples the covalently bonded zero-layer, restoring quasi-free-standing graphene with excellent charge neutrality, while simultaneously preserving long-range structural coherence, homogeneity, and intrinsic band dispersion. Raman mapping reveals distinct intercalation pathways and demonstrates that diffusion-mediated intercalation suppresses defect formation, enabling superior graphene quality with homogeneous strain and charge carrier distribution. Temperature-dependent diffraction and vibrational analysis further indicate proximity-induced strain coupling between graphene and the commensurate triangular Sn(1$\times$1) interface, highlighting the role of interfacial confinement in tailoring lattice and electronic interactions. Beyond this specific material system, our results provide a framework for engineering buried two-dimensional metal layers beneath graphene and other van der Waals materials, opening opportunities for controlled metal-graphene heterostructures with tunable interfacial coupling and functionality.

%---------------------------------------------------------------------------------------
\section*{acknowledgments}
This work was supported by the German Research Foundation (Deutsche Forschungsgemeinschaft, DFG) within the Research Unit FOR5242 (Project 449119662) and an individual research grant (Project 509747664).
\section*{Author Contributions} Z.M. conceived and designed the experiments, prepared the samples, performed the experiments, analyzed the data, and wrote the manuscript with significant input from all authors. N.T. and N.B. contributed to experiments and data analysis. T.S., D.R.T.Z., and C.T. contributed to the conception of the project. 
All authors reviewed and approved the manuscript.

%************************************************************************
\bibliography{GrapheneLibrary21022025}

@article{Mamiyev2025a,
  title={Enhanced Light--Matter Interactions With a Single {Sn} Nanoantenna on Epitaxial Graphene},
  author={Mamiyev, Zamin and Balayeva, Narmina O and Zahn, Dietrich RT and Tegenkamp, Christoph},
  journal={Advanced Optical Materials},
  volume={13},
  number={36},
  pages={e00979},
  year={2025},
  doi = {10.1002/adom.202500979},
  publisher={Wiley Online Library}
}

@article{Harling2025,
  title={Mesoscopic Scale Study of Lateral Dynamics of Sn Intercalation of the Graphene Buffer Layer on SiC},
  author={Harling, Benno and Mamiyev, Zamin and Tegenkamp, Christoph and Wenderoth, Martin},
  journal={Carbon},
  pages={120711},
  year={2025},
  doi = {10.1016/j.carbon.2025.120711},
  publisher={Elsevier}
}

@article{Mamiyev2025,
  title={Confinement induced strain effects in epitaxial graphene},
  author={Mamiyev, Zamin and Balayeva, Narmina O and Ghosal, Chitran and Zahn, Dietrich RT and Tegenkamp, Christoph},
  journal={Carbon},
  volume={234},
  pages={120002},
  year={2025},
   doi = {10.1016/j.carbon.2025.120002},
  publisher={Elsevier}
}

@article{Thoai2025,
  title={Intercalant-induced Kekule ordering and gap opening in quasi-free-standing graphene},
  author={Huu Thoai Ngo and Zamin Mamiyev and Niklas Witt and Tim Wehling and Christoph Tegenkamp},
  journal={arXiv preprint arXiv:2512.20366},
  year={2025},
 doi = {10.48550/arXiv.2512.20366}
}

@article{Mamiyev2024,
  title={Exploring graphene-substrate interactions: plasmonic excitation in Sn-intercalated epitaxial graphene},
  author={Mamiyev, Zamin and Tegenkamp, Christoph},
  journal={2D Materials},
  volume={11},
  number={2},
  pages={025013},
  year={2024},
 doi={10.1088/2053-1583/ad1a70},
  publisher={IOP Publishing}
}

@article{Schadlich2023,
  title={Domain Boundary Formation Within an Intercalated {Pb} Monolayer Featuring Charge-Neutral Epitaxial Graphene},
  author={Sch{\"a}dlich, Philip and Ghosal, Chitran and Stettner, Monja and Matta, Bharti and Wolff, Susanne and Sch{\"o}lzel, Franziska and Richter, Peter and Hutter, Mark and Haags, Anja and Wenzel, Sabine and others},
  journal={Advanced Materials Interfaces},
  volume={10},
  number={27},
  pages={2300471},
  year={2023},
  doi = {10.1002/admi.202300471},
  publisher={Wiley Online Library}
}

@article{Mamiyev2022,
  title={Sn intercalation into the {BL/SiC (0001)} interface: A detailed {SPA-LEED} investigation},
  author={Mamiyev, Zamin and Tegenkamp, Christoph},
  journal={Surfaces and Interfaces},
  volume={34},
  pages={102304},
  year={2022},
  doi = {10.1016/j.surfin.2022.102304},
  publisher={Elsevier}
}

@article{Wundrack2025,
  title={Lithographically Controlled Liquid Metal Diffusion in Graphene: Fabrication and Magnetotransport Signatures of Superconductivity},
  author={Wundrack, Stefan and Bothe, Marc and Jaime, Marcelo and K{\"u}ster, Kathrin and Gruschwitz, Markus and Yin, Yefei and Mamiyev, Zamin and Sch{\"a}dlich, Philip and Matta, Bharti and Datta, Sawani and others},
  journal={Advanced Materials},
  volume={38},
  number={5},
  pages={e11992},
  year={2026},
  doi = {10.1002/adma.202511992},
  publisher={Wiley Online Library}
}

@article{Tilgner2025,
  title={{Si intercalation beneath epitaxial graphene: screening Mott states at the SiC(0001) interface}},
  author={Tilgner, Niclas and Mamiyev, Zamin and Wolff, Susanne and Sch{\"a}dlich, Philip and G{\"o}hler, Fabian and Tegenkamp, Christoph and Seyller, Thomas},
  journal={2D Mater.},
  volume={12},
  number={4},
  pages={045022},
  year={2025},
  publisher={IOP Publishing},
  doi={10.1088/2053-1583/ae0f26}
}

@article{Ghosal2025,
  title = {Mott states proximitized to a relativistic electron gas in epitaxial graphene},
  author = {Ghosal, Chitran and Ryee, Siheon and Mamiyev, Zamin and Federl, Maria-Elisabeth and Witt, Niklas and Wehling, Tim and Tegenkamp, Christoph},
  journal = {Phys. Rev. B},
  volume = {111},
  issue = {23},
  pages = {235426},
  numpages = {12},
  year = {2025},
  month = {Jun},
  publisher = {American Physical Society},
  doi = {10.1103/vk9s-zzcw}
}

@article{Gruschwitz2025,
  title={{From stripes to hexagons: strain-induced 2D Pb phases confined between graphene and SiC}},
  author={Gruschwitz, Markus and Sologub, Sergii and Ghosal, Chitran and Mamiyev, Zamin and Niu, Yuran and Zakharov, Alexei and Tegenkamp, Christoph},
  journal={Advanced Materials Interfaces},
  volume={12},
  number={21},
  pages={e00617},
  year={2025},
  doi = {10.1002/admi.202500617},
  publisher={Wiley Online Library}
}

@article{Lindsay2010,
  title = {Flexural phonons and thermal transport in graphene},
  author = {Lindsay, L. and Broido, D. A. and Mingo, Natalio},
  journal = {Phys. Rev. B},
  volume = {82},
  issue = {11},
  pages = {115427},
  numpages = {6},
  year = {2010},
  month = {Sep},
  publisher = {American Physical Society},
  doi = {10.1103/PhysRevB.82.115427}
}

@article{Das2008,
  title={Monitoring dopants by Raman scattering in an electrochemically top-gated graphene transistor},
  author={Das, Anindya and Pisana, Simone and Chakraborty, Biswanath and Piscanec, Stefano and Saha, Srijan K and Waghmare, Umesh V and Novoselov, Konstantin S and Krishnamurthy, Hulikal R and Geim, Andre K and Ferrari, Andrea C and others},
  journal={Nature nanotechnology},
  volume={3},
  number={4},
  pages={210--215},
  year={2008},
  doi = {10.1038/nnano.2008.67},
  publisher={Nature Publishing Group UK London}
}

@article{Cheng2011,
  title = {{Gr\"uneisen} parameter of the {G} mode of strained monolayer graphene},
  author = {Cheng, Y. C. and Zhu, Z. Y. and Huang, G. S. and Schwingenschl\"ogl, U.},
  journal = {Phys. Rev. B},
  volume = {83},
  issue = {11},
  pages = {115449},
  numpages = {5},
  year = {2011},
  month = {Mar},
  publisher = {American Physical Society},
  doi = {10.1103/PhysRevB.83.115449}
}

@article{Mohiuddin2009,
  title = {Uniaxial strain in graphene by {Raman} spectroscopy: {$G$} peak splitting, {Gr\"uneisen} parameters, and sample orientation},
  author = {Mohiuddin, T. M. G. and Lombardo, A. and Nair, R. R. and Bonetti, A. and Savini, G. and Jalil, R. and Bonini, N. and Basko, D. M. and Galiotis, C. and Marzari, N. and Novoselov, K. S. and Geim, A. K. and Ferrari, A. C.},
  journal = {Phys. Rev. B},
  volume = {79},
  issue = {20},
  pages = {205433},
  numpages = {8},
  year = {2009},
  month = {May},
  publisher = {American Physical Society},
  doi = {10.1103/PhysRevB.79.205433}
}

@article{Venezuela2011,
  title = {Theory of double-resonant {Raman spectra} in graphene: Intensity and line shape of defect-induced and two-phonon bands},
  author = {Venezuela, Pedro and Lazzeri, Michele and Mauri, Francesco},
  journal = {Phys. Rev. B},
  volume = {84},
  issue = {3},
  pages = {035433},
  numpages = {25},
  year = {2011},
  month = {Jul},
  publisher = {American Physical Society},
  doi = {10.1103/PhysRevB.84.035433}
}

@article{Attaccalite2010,
  title={Doped graphene as tunable electron- phonon coupling material},
  author={Attaccalite, Claudio and Wirtz, Ludger and Lazzeri, Michele and Mauri, Francesco and Rubio, Angel},
  journal={Nano letters},
  volume={10},
  number={4},
  pages={1172--1176},
  year={2010},
  doi = {10.1021/nl9034626},
  publisher={ACS Publications}
}

@article{Basko2009,
  title = {Electron-electron interactions and doping dependence of the two-phonon {Raman} intensity in graphene},
  author = {Basko, D. M. and Piscanec, S. and Ferrari, A. C.},
  journal = {Phys. Rev. B},
  volume = {80},
  issue = {16},
  pages = {165413},
  numpages = {10},
  year = {2009},
  month = {Oct},
  publisher = {American Physical Society},
  doi = {10.1103/PhysRevB.80.165413}
}

@article{Mueller2017,
  title={Evaluating arbitrary strain configurations and doping in graphene with {Raman} spectroscopy},
  author={Mueller, Niclas S and Heeg, Sebastian and Alvarez, Miriam Pe{\~n}a and Kusch, Patryk and Wasserroth, S{\"o}ren and Clark, Nick and Schedin, Fredrik and Parthenios, John and Papagelis, Konstantinos and Galiotis, Costas and others},
  journal={2D Materials},
  volume={5},
  number={1},
  pages={015016},
  year={2017},
  doi = {10.1088/2053-1583/aa90b3},
  publisher={IOP Publishing}
}

@article{Lee2012,
  title={Optical separation of mechanical strain from charge doping in graphene},
  author={Lee, Ji Eun and Ahn, Gwanghyun and Shim, Jihye and Lee, Young Sik and Ryu, Sunmin},
  journal={Nature communications},
  volume={3},
  number={1},
  pages={1024},
  year={2012}, 
  doi = {10.1038/ncomms2022},
  publisher={Nature Publishing Group UK London}
}

@article{Canccado2011,
  title={Quantifying defects in graphene via {Raman} spectroscopy at different excitation energies},
  author={Can{\c{c}}ado, L Gustavo and Jorio, Ado and Ferreira, EH Martins and Stavale, F and Achete, Carlos Alberto and Capaz, Rodrigo Barbosa and Moutinho, MV de O and Lombardo, Antonio and Kulmala, TS and Ferrari, Andrea Carlo},
  journal={Nano letters},
  volume={11},
  number={8},
  pages={3190--3196},
  year={2011},
  doi = {10.1021/nl201432g},
  publisher={ACS Publications}
}

@article{Turker2023,
  title={{2D oxides} realized via confinement heteroepitaxy},
  author={Turker, Furkan and Dong, Chengye and Wetherington, Maxwell T and El-Sherif, Hesham and Holoviak, Stephen and Trdinich, Zachary J and Lawson, Eric T and Krishnan, Gopi and Whittier, Caleb and Sinnott, Susan B and others},
  journal={Advanced Functional Materials},
  volume={33},
  number={5},
  pages={2210404},
  year={2023},
  doi = {10.1002/smll.202505640},
  publisher={Wiley Online Library}
}

@article{Lucchese2010,
  title={Quantifying ion-induced defects and {Raman} relaxation length in graphene},
  author={Lucchese, M{\'a}rcia Maria and Stavale, F and Ferreira, EH Martins and Vilani, Cec{\i}lia and Moutinho, Marcus Vinicius de Oliveira and Capaz, Rodrigo B and Achete, Carlos Alberto and Jorio, A},
  journal={Carbon},
  volume={48},
  number={5},
  pages={1592--1597},
  year={2010},
  doi = {10.1016/j.carbon.2009.12.057},
  publisher={Elsevier}
}

@article{Ferrari2007,
  title={Raman spectroscopy of graphene and graphite: Disorder, electron--phonon coupling, doping and nonadiabatic effects},
  author={Ferrari, Andrea C},
  journal={Solid state communications},
  volume={143},
  number={1-2},
  pages={47--57},
  year={2007},
  doi = {10.1016/j.ssc.2007.03.052},
  publisher={Elsevier}
}

@article{Eckmann2012,
  title={Probing the nature of defects in graphene by {Raman} spectroscopy},
  author={Eckmann, Axel and Felten, Alexandre and Mishchenko, Artem and Britnell, Liam and Krupke, Ralph and Novoselov, Kostya S and Casiraghi, Cinzia},
  journal={Nano letters},
  volume={12},
  number={8},
  pages={3925--3930},
  year={2012},
  doi = {10.1021/nl300901a},
  publisher={ACS Publications}
}

@article{Pompei2025,
  title={Novel structures of Gallenene intercalated in epitaxial Graphene},
  author={Pompei, Emanuele and Skibi{\'n}ska, Katarzyna and Senesi, Giulio and Vlamidis, Ylea and Rossi, Antonio and Forti, Stiven and Coletti, Camilla and Beltram, Fabio and Rubini, Silvia and Sorba, Lucia and others},
  journal={Small},
  volume={21},
  number={38},
  pages={e05640},
  year={2025},
  doi = {10.1002/smll.202505640},
  publisher={Wiley Online Library}
}

@article{Schumann2014,
  title = {Effect of buffer layer coupling on the lattice parameter of epitaxial graphene on {SiC(0001)}},
  author = {Schumann, T. and Dubslaff, M. and Oliveira, M. H. and Hanke, M. and Lopes, J. M. J. and Riechert, H.},
  journal = {Phys. Rev. B},
  volume = {90},
  issue = {4},
  pages = {041403},
  numpages = {5},
  year = {2014},
  month = {Jul},
  publisher = {American Physical Society},
  doi = {10.1103/PhysRevB.90.041403}
}

@article{Li1986,
  title={Thermal expansion of the hexagonal {(4H) polytype of SiC}},
  author={Li, Z and Bradt, Richard C},
  journal={Journal of applied physics},
  volume={60},
  number={2},
  pages={612--614},
  year={1986}, 
  doi = {10.1063/1.337456},
  publisher={American Institute of Physics}
}

@article{Ferralis2008,
  title = {Evidence of Structural Strain in Epitaxial Graphene Layers on {6H-SiC(0001)}},
  author = {Ferralis, Nicola and Maboudian, Roya and Carraro, Carlo},
  journal = {Phys. Rev. Lett.},
  volume = {101},
  issue = {15},
  pages = {156801},
  numpages = {4},
  year = {2008},
  month = {Oct},
  publisher = {American Physical Society},
  doi = {10.1103/PhysRevLett.101.156801},
  url = {https://link.aps.org/doi/10.1103/PhysRevLett.101.156801}
}

@article{Chen2019,
  title = {Diffraction paradox: An unusually broad diffraction background marks high quality graphene},
  author = {Chen, S. and Horn von Hoegen, M. and Thiel, P. A. and Tringides, M. C.},
  journal = {Phys. Rev. B},
  volume = {100},
  issue = {15},
  pages = {155307},
  numpages = {6},
  year = {2019},
  month = {Oct},
  publisher = {American Physical Society},
  doi = {10.1103/PhysRevB.100.155307}
}

@Article{Hoegen1999,
  Title                    = {{Growth of semiconductor layers studied by spot profile analysing low energy electron diffraction}},
  Author                   = {{Horn-von Hoegen}, Michael},
  Journal                  = {Zeitschrift f{\"{u}}r Krist.},
  Year                     = {1999},
  Number                   = {10},
  Pages                    = {591--629},
  doi = {/10.1524/zkri.1999.214.10.591},
  Volume                   = {214},
  Publisher                = {Oldenbourg},
  Url                      = {http://cat.inist.fr/?aModele=afficheN{\&}cpsidt=1959726}
}

@article{Fasolino2007,
  title={Intrinsic ripples in graphene},
  author={Fasolino, Annalisa and Los, JH and Katsnelson, Mikhail I},
  journal={Nature materials},
  volume={6},
  number={11},
  pages={858--861},
  year={2007},
  doi = {10.1038/nmat2011},
  publisher={Nature Publishing Group UK London}
}

@article{Ichinokura2016,
  title={{Superconducting calcium-intercalated bilayer graphene}},
  author={Ichinokura, Satoru and Sugawara, Katsuaki and Takayama, Akari and Takahashi, Takashi and Hasegawa, Shuji},
  journal={ACS nano},
  volume={10},
  number={2},
  pages={2761--2765},
  year={2016},
  doi = {10.1021/acsnano.5b07848},
  publisher={ACS Publications}
}

@article{Ristein2012,
  title = {Origin of Doping in Quasi-Free-Standing Graphene on Silicon Carbide},
  author = {Ristein, J. and Mammadov, S. and Seyller, Th.},
  journal = {Phys. Rev. Lett.},
  volume = {108},
  issue = {24},
  pages = {246104},
  numpages = {5},
  year = {2012},
  month = {Jun},
  publisher = {American Physical Society},
  doi = {10.1103/PhysRevLett.108.246104}
}

@article{qteish1992electronic,
  title={{Electronic-charge displacement around a stacking boundary in SiC polytypes}},
  author={Qteish, A and Heine, Volker and Needs, R J},
  journal={Phys. Rev. B},
  volume={45},
  number={12},
  pages={6376},
  year={1992},
  publisher={APS},
  doi={10.1103/PhysRevB.45.6376}
}

@article{Novoselov2016,
  title={2D materials and van der Waals heterostructures},
  author={Novoselov, K S and Mishchenko, Artem and Carvalho, Alexandra and Castro Neto, AH},
  journal={Science},
  volume={353},
  number={6298},
  pages={aac9439},
  year={2016},
  doi = {10.1126/science.aac9439},
  publisher={American Association for the Advancement of Science}
}

@article{Malard2009,
  title={Raman spectroscopy in graphene},
  author={Malard, Leandro M and Pimenta, Marcos Assun{\c{c}}{\~a}o and Dresselhaus, Gene and Dresselhaus, Mildred Spiewak},
  journal={Physics reports},
  volume={473},
  number={5-6},
  pages={51--87},
  year={2009},
  doi = {10.1016/j.physrep.2009.02.003},
  publisher={Elsevier}
}

@article{Ferrari2013,
  title={Raman spectroscopy as a versatile tool for studying the properties of graphene},
  author={Ferrari, Andrea C and Basko, Denis M},
  journal={Nature nanotechnology},
  volume={8},
  number={4},
  pages={235--246},
  year={2013},
  doi = {10.1038/nnano.2013.46},
  publisher={Nature Publishing Group UK London}
}

@article{de2007epitaxial,
  title={Epitaxial graphene},
  author={De Heer, Walt A and Berger, Claire and Wu, Xiaosong and First, Phillip N and Conrad, Edward H and Li, Xuebin and Li, Tianbo and Sprinkle, Michael and Hass, Joanna and Sadowski, Marcin L and others},
  journal={Solid State Commun.},
  volume={143},
  number={1-2},
  pages={92--100},
  year={2007},
  publisher={Elsevier},
  doi = {10.1016/j.ssc.2007.04.023}
}

@article{Briggs2020,
  title={Atomically thin half-van der Waals metals enabled by confinement heteroepitaxy},
  author={Briggs, Natalie and Bersch, Brian and Wang, Yuanxi and Jiang, Jue and Koch, Roland J and Nayir, Nadire and Wang, Ke and Kolmer, Marek and Ko, Wonhee and De La Fuente Duran, Ana and others},
  journal={Nat. Mater.},
  volume={19},
  number={6},
  pages={637--643},
  year={2020},
  publisher={Nature Publishing Group UK London},
  doi = {10.1038/s41563-020-0631-x}
}

@article{Mamiyev2021a,
  title = {Enforced Long-Range Order in {1D} Wires by Coupling to Higher Dimensions},
  author = {Mamiyev, Zamin and Fink, Christa and Holtgrewe, Kris and Pfn\"ur, Herbert and Sanna, Simone},
  journal = {Phys. Rev. Lett.},
  volume = {126},
  issue = {10},
  pages = {106101},
  numpages = {6},
  year = {2021},
  month = {Mar},
  publisher = {American Physical Society},
  doi = {10.1103/PhysRevLett.126.106101}
}

@article{Scheithauer1986,
  title={A new LEED instrument for quantitative spot profile analysis},
  author={Scheithauer, U and Meyer, G and Henzler, M},
  journal={Surface Science},
  volume={178},
  number={1-3},
  pages={441--451},
  year={1986},
  doi = {10.1016/0039-6028(86)90321-3},
  publisher={Elsevier}
}

@article{Rohrl2008,
  title={Raman spectra of epitaxial graphene on {SiC(0001)}},
  author={R{\"o}hrl, J and Hundhausen, M and Emtsev, KV and Seyller, Th and Graupner, R and Ley, LJAPL},
  journal={Applied Physics Letters},
  volume={92},
  number={20},
  year={2008}, 
  doi = {10.1063/1.2929746},
  publisher={AIP Publishing}
}

@article{Ferralis2011,
  title = {Determination of substrate pinning in epitaxial and supported graphene layers via {Raman} scattering},
  author = {Ferralis, Nicola and Maboudian, Roya and Carraro, Carlo},
  journal = {Phys. Rev. B},
  volume = {83},
  issue = {8},
  pages = {081410},
  numpages = {4},
  year = {2011},
  month = {Feb},
  publisher = {American Physical Society},
  doi = {10.1103/PhysRevB.83.081410}
}

@inproceedings{Kestle2000,
  title={A {UHV} study of {Ni/SiC} Schottky barrier and ohmic contact formation},
  author={Kestle, A and Wilks, SP and Dunstan, PR and Pritchard, M and Pope, G and Koh, A and Mawby, Philip Andrew},
  booktitle={Materials Science Forum},
  volume={338},
  pages={1025--1028},
  year={2000}, 
  doi = {10.4028/www.scientific.net/MSF.338-342.1025},
  organization={Trans Tech Publ}
}

@article{Zhao2011,
  title={Intercalation of few-layer graphite flakes with FeCl3: Raman determination of Fermi level, layer by layer decoupling, and stability},
  author={Zhao, Weijie and Tan, Ping Heng and Liu, Jian and Ferrari, Andrea C},
  journal={Journal of the American Chemical Society},
  volume={133},
  number={15},
  pages={5941--5946},
  year={2011},
  doi = {10.1021/ja110939a},
  publisher={ACS Publications}
}

@article{Emtsev2008,
  title = {Interaction, growth, and ordering of epitaxial graphene on SiC{0001} surfaces: A comparative photoelectron spectroscopy study},
  author = {Emtsev, K. V. and Speck, F. and Seyller, Th. and Ley, L. and Riley, J. D.},
  journal = {Phys. Rev. B},
  volume = {77},
  issue = {15},
  pages = {155303},
  numpages = {10},
  year = {2008},
  month = {Apr},
  publisher = {American Physical Society},
  doi = {10.1103/PhysRevB.77.155303}
}

@article{Lazzeri2006,
  title = {Nonadiabatic Kohn Anomaly in a Doped Graphene Monolayer},
  author = {Lazzeri, Michele and Mauri, Francesco},
  journal = {Phys. Rev. Lett.},
  volume = {97},
  issue = {26},
  pages = {266407},
  numpages = {4},
  year = {2006},
  month = {Dec},
  publisher = {American Physical Society},
  doi = {10.1103/PhysRevLett.97.266407}
}

@article{Yoon2011,
  title = {Strain-Dependent Splitting of the Double-Resonance {Raman} Scattering Band in Graphene},
  author = {Yoon, Duhee and Son, Young-Woo and Cheong, Hyeonsik},
  journal = {Phys. Rev. Lett.},
  volume = {106},
  issue = {15},
  pages = {155502},
  numpages = {4},
  year = {2011},
  month = {Apr},
  publisher = {American Physical Society},
  doi = {10.1103/PhysRevLett.106.155502}
}

@article{Yoon2011a,
  title={Negative thermal expansion coefficient of graphene measured by Raman spectroscopy},
  author={Yoon, Duhee and Son, Young-Woo and Cheong, Hyeonsik},
  journal={Nano letters},
  volume={11},
  number={8},
  pages={3227--3231},
  year={2011},
  doi = {10.1021/nl201488g},
  publisher={ACS Publications}
}

@article{Axdal1987,
  title={A theory for the kinetics of intercalation of graphite},
  author={Axdal, SH Anderson and Chung, DDL},
  journal={Carbon},
  volume={25},
  number={3},
  pages={377--389},
  year={1987},
  doi = {10.1016/0008-6223(87)90009-1},
  publisher={Elsevier}
}

@article{Schmitt2024,
  title={Achieving environmental stability in an atomically thin quantum spin Hall insulator via graphene intercalation},
  author={Schmitt, Cedric and Erhardt, Jonas and Eck, Philipp and Schmitt, Matthias and Lee, Kyungchan and Ke{\ss}ler, Philipp and Wagner, Tim and Spring, Merit and Liu, Bing and Enzner, Stefan and others},
  journal={Nature Communications},
  volume={15},
  number={1},
  pages={1486},
  year={2024},
  doi = {10.1038/s41467-024-45816-9},
  publisher={Nature Publishing Group UK London}
}

@article{Vera2024,
  title={{Large-area intercalated two-dimensional pb/graphene heterostructure as a platform for generating spin--orbit torque}},
  author={Vera, Alexander and Zheng, Boyang and Yanez, Wilson and Yang, Kaijie and Kim, Seong Yeoul and Wang, Xinglu and Kotsakidis, Jimmy C and El-Sherif, Hesham and Krishnan, Gopi and Koch, Roland J and others},
  journal={ACS nano},
  volume={18},
  number={33},
  pages={21985--21997},
  year={2024},
  doi = {10.1021/acsnano.4c04075},
  publisher={ACS Publications}
}

@article{Emtsev2009,
  title={Towards wafer-size graphene layers by atmospheric pressure graphitization of silicon carbide},
  author={Emtsev, Konstantin V and Bostwick, Aaron and Horn, Karsten and Jobst, Johannes and Kellogg, Gary L and Ley, Lothar and McChesney, Jessica L and Ohta, Taisuke and Reshanov, Sergey A and R{\"o}hrl, Jonas and others},
  journal={Nat. Mater.},
  volume={8},
  number={3},
  pages={203--207},
  year={2009},
  doi = {10.1038/nmat2382},
  publisher={Nature Publishing Group}
}

@article{Hayashi2017,
  title={{Triangular lattice atomic layer of Sn(1$\times$1) at graphene/SiC (0001) interface}},
  author={Hayashi, Shingo and Visikovskiy, Anton and Kajiwara, Takashi and Iimori, Takushi and Shirasawa, Tetsuroh and Nakastuji, Kan and Miyamachi, Toshio and Nakashima, Shuhei and Yaji, Koichiro and Mase, Kazuhiko and others},
  journal={Appl. Phys. Express},
  volume={11},
  number={1},
  pages={015202},
  year={2017},
  publisher={IOP Publishing},
  doi = {10.7567/APEX.11.015202}
}

@article{Visikovskiy2018,
  title={{Computational study of heavy group IV elements (Ge, Sn, Pb) triangular lattice atomic layers on SiC (0001) surface}},
  author={Visikovskiy, Anton and Hayashi, Shingo and Kajiwara, Takashi and Komori, Fumio and Yaji, Koichiro and Tanaka, Satoru},
  journal={arXiv preprint arXiv:1809.00829},
  year={2018},
  doi = {10.48550/arXiv.1809.00829}
}

@article{Ferrari2015,
  title={Science and technology roadmap for graphene, related two-dimensional crystals, and hybrid systems},
  author={Ferrari, Andrea C and Bonaccorso, Francesco and Fal'Ko, Vladimir and Novoselov, Konstantin S and Roche, Stephan and B{\o}ggild, Peter and Borini, Stefano and Koppens, Frank HL and Palermo, Vincenzo and Pugno, Nicola and others},
  journal={Nanoscale},
  volume={7},
  number={11},
  pages={4598--4810},
  year={2015},
  doi = {10.1039/C4NR01600A},
  publisher={Royal Society of Chemistry}
}

@article{Ares2022,
  title={Recent advances in graphene and other {2D materials}},
  author={Ares, Pablo and Novoselov, Kostya S},
  journal={Nano Materials Science},
  volume={4},
  number={1},
  pages={3--9},
  year={2022},
  doi = {10.1016/j.nanoms.2021.05.002},
  publisher={Elsevier}
}
%************************************************************************

\end{document}